\pdfoutput=1  
\documentclass[a4paper,11pt]{article}
\usepackage{jheppub}
\usepackage[utf8]{inputenc}
\usepackage{graphicx,amsmath,amsfonts,amssymb,slashed}
\usepackage{numprint}
\usepackage{enumitem}
 \usepackage{mathtools}        
\usepackage{graphicx,color,amsmath}
\usepackage{hyperref}
\usepackage{soul}
\usepackage{placeins}
\usepackage{booktabs, tabularx, makecell, multirow, xspace}

\newcolumntype{C}[1]{>{\centering\let\newline\\\arraybackslash\hspace{0pt}}m{#1}}
\newcolumntype{L}[1]{>{\raggedright\let\newline\\\arraybackslash\hspace{0pt}}m{#1}}

\allowdisplaybreaks

\usepackage{bm}
\usepackage{mathtools}

\graphicspath{{figures/}}

\usepackage[dvipsnames]{xcolor}
\definecolor{red}{rgb}{0.9, 0,0}
\definecolor{cerulean}{rgb}{0., 0.42,0.9}

\title{The Postdoc Accord in Theoretical High Energy Physics}

\author[a]{Djuna Croon,}
\author[b]{Patrick J. Fox,}
\author[b]{Roni Harnik,}
\author[c,d]{Simon Knapen,}
\author[e, f]{Mariangela Lisanti,}
\author[g]{Lina Necib,}
\author[h]{and Tien-Tien Yu}

\affiliation[a]{\normalsize\it Institute for Particle Physics Phenomenology, Department of Physics, Durham University, Durham DH1 3LE, U.K.}
\affiliation[b]{\normalsize\it Theory Division, Fermilab, Batavia, IL 60510, USA}
\affiliation[c]{Theoretical Physics Group, Lawrence Berkeley National Laboratory, Berkeley, CA 94720, USA}
\affiliation[d]{Berkeley Center for Theoretical Physics, Department of Physics, University of California, Berkeley, CA 94720, USA}
\affiliation[e]{Department of Physics, Princeton University, Princeton, NJ 08544, USA}
\affiliation[f]{Center for Computational Astrophysics, Flatiron Institute, New York, NY 10010, USA} 
\affiliation[g]{\normalsize\it Department of Physics, Massachusetts Institute of Technology, Cambridge, MA 02139, USA}
\affiliation[h]{\normalsize\it Department of Physics and Institute for Fundamental Science,
University of Oregon, Eugene, Oregon 97403}
\emailAdd{het.postdoc.deadline@gmail.com}

\abstract{
We present the results of a survey meant to assess the opinion of the high-energy physics theory (HET) community on the January~7th postdoc acceptance deadline---specifically, whether there is a preference to shift the deadline to later in January or February.  This survey, which served for information-gathering purpose only, is part of a community conversation on the optimal timing of an acceptance deadline and whether the community would be better served by a later date.  In addition, we present an analysis of data from the postdoc Rumor Mill, which gives a picture of the current hiring landscape in the field. We emphasize the importance of preserving a universal deadline, and the current results of our survey show broad support for a shift to a later date. A link to the survey, frequently asked questions, a running list of supporters, and next steps can be found \href{https://het-postdoc-accord.github.io/information/}{here}.
} 

\begin{document}

\maketitle

\section{\label{sec:intro}Introduction} 

Since 2007, the majority of the high-energy physics theory (HET) community has adopted an agreement in which most institutions have pledged to set postdoc acceptance deadlines no earlier than January~7th~\cite{Jan7}. This agreement has been critical for establishing a standard of equity and fairness in hiring and recruiting practices, allowing several generations of postdocs to make life and career decisions with the maximum amount of information at their disposal. 

Through our own experiences, as well as interactions with community members, several challenges associated with the specific choice of January~7th as the deadline have become apparent. Specifically, the weeks between December 25th and January 7th are highly stressful for postdocs still waiting for offers. In much of the world, this timing coincides with the Christmas \& New Year holiday period, during which many are away from work. This means that once an applicant receives an offer, they have more difficulty getting in touch with their senior colleagues to solicit advice on their options. These issues are further compounded for applicants who balance family commitments like child-care and elder-care during the holiday season. 

On the hiring side, the lack of administrative support during this time can make it more difficult to respond promptly as initial offers are being declined, or if applicants have questions which require input from Human Resources (HR). The holiday break, moreover, means that colleagues are often scattered across the globe, making it difficult to find enough time to have thoughtful discussions about the candidates, especially for second and third round offers, which often need to happen on short notice. 

The January 7th deadline also increasingly conflicts with established deadlines in the astroparticle and cosmology communities, which have a growing overlap with the HET community. The American Astronomical Society’s (AAS) policy, adopted in 1988 and later reaffirmed in 2003 and 2006, sets the common acceptance deadline to February~15th~\cite{AAS}, which means that astroparticle applicants often have to make important career decisions with incomplete information. 
Similarly, the mathematics community has also reached an agreement with a deadline of February 6th~\cite{AMS}.

To gather a broader picture on how our community at large views these issues, we conducted an online survey. We here report on results of this survey, which collected data from May 3 to June 13, 2023, and briefly comment on next steps.

This note is organized as follows. In Sec.~\ref{sec:rumormill}, we present the results of an analysis of the high-energy physics~(HEP) theory Rumor Mill data, which shows when offers are typically extended, accepted and declined, and for how long candidates typically consider an offer. In Sec.~\ref{sec:survey}, we describe the results of the survey, and then we summarize some of the write-in feedback in Sec.~\ref{sec:comments}. For the latter section, we  paraphrase all responses, to ensure individual respondents cannot be identified. We conclude in Sec.~\ref{sec:conclusions}. The raw survey data, with all identifiable information and write-in comments redacted, is publicly available for others to analyse~\cite{website}.

\section{Postdoc Rumor Mill Data Analysis\label{sec:rumormill}}
To understand the current state of affairs, we analysed the data from the postdoc Rumor Mill from the last six years (2018 to 2023) \cite{rumormill}. We found no significant differences between the datasets on a year-by-year basis, even during the COVID era, and therefore group all years together to increase the statistics of the sample. We state up front that this data is inherently incomplete, as not all applicants post to the Rumor Mill, or only post partial information. We point out the instances where this may bias the reading of the results.

\begin{figure}\centering
\includegraphics[width=\textwidth]{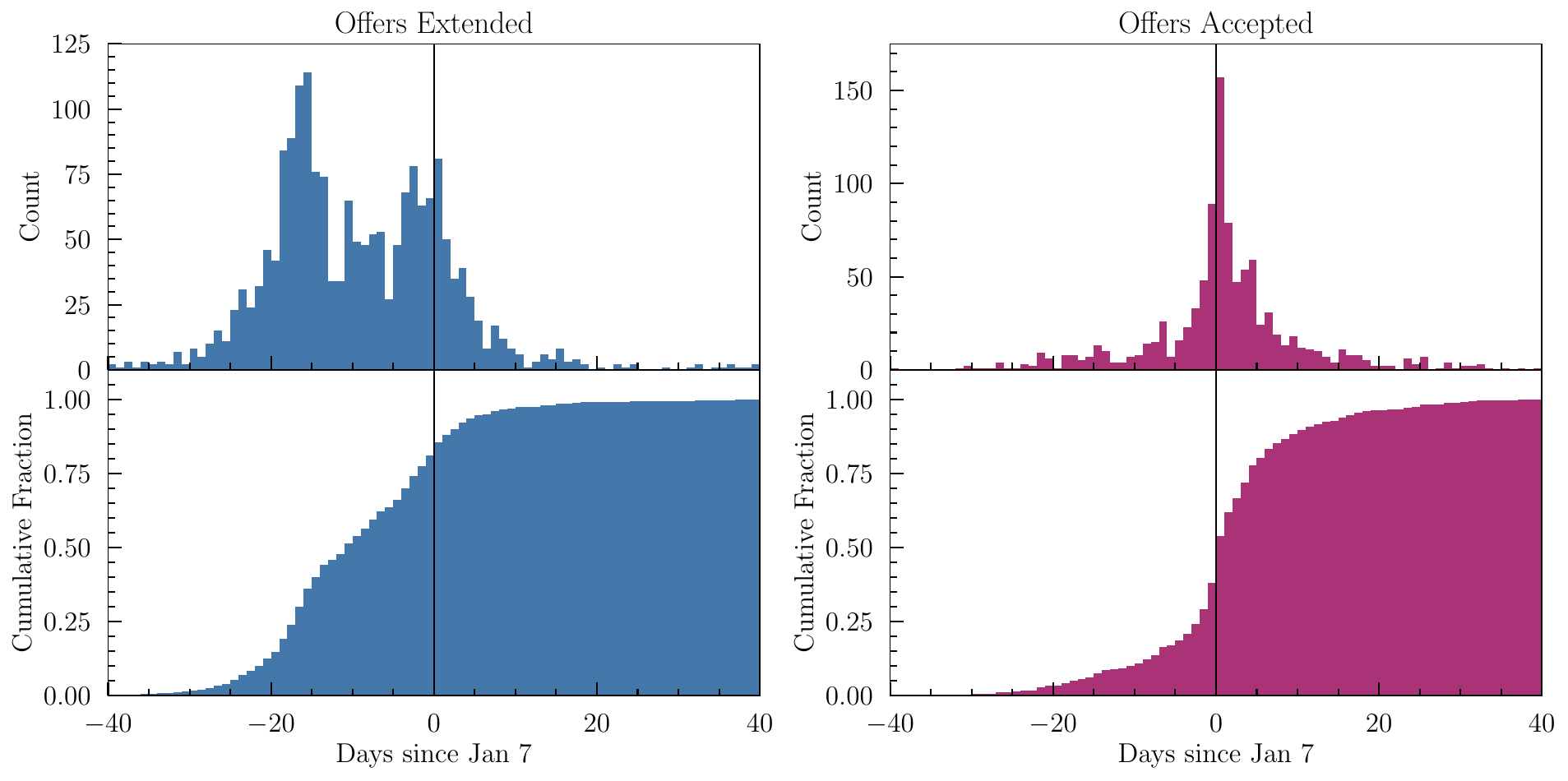}
\caption{Distributions~\textbf{(top row)} and cumulative distributions~\textbf{(bottom row)} of the dates at which offers were extended~\textbf{(left column)} and accepted~\textbf{(right column)}, relative to January 7th, for the years 2018 to 2023. 
The total sample size for the offers is 1906; the total sample size for the accepted offers is 1025.\label{jan7plot}}
\end{figure}

The distributions of the offers extended and offers accepted are shown in Fig.~\ref{jan7plot}. Most offers are extended about two weeks before January 7th, with a significant second peak on and right after January 7th. As expected, there is a large peak in offers accepted on January 7th. Curiously, there is also a minor but significant peak on January 1st. One must keep in mind,  however, that there is often a delay between the acceptance of an offer and it appearing on the Rumor Mill, which introduces a systematic bias. The cumulative distributions reveal that the market converges relatively quickly after January~7th. Specifically, the 90\% and 95\% quantiles of the ``offers accepted'' distribution are at 13.6 and 24.8 days after January 7th, respectively. However, it is plausible that applicants receiving and accepting offers well after January 7th are less likely to post to the Rumor Mill. On the other hand, the expected delay between accepting an offer and having it appear on the Rumor Mill would skew the true distribution towards faster convergence.

\begin{figure}\centering
\includegraphics[width=\textwidth]{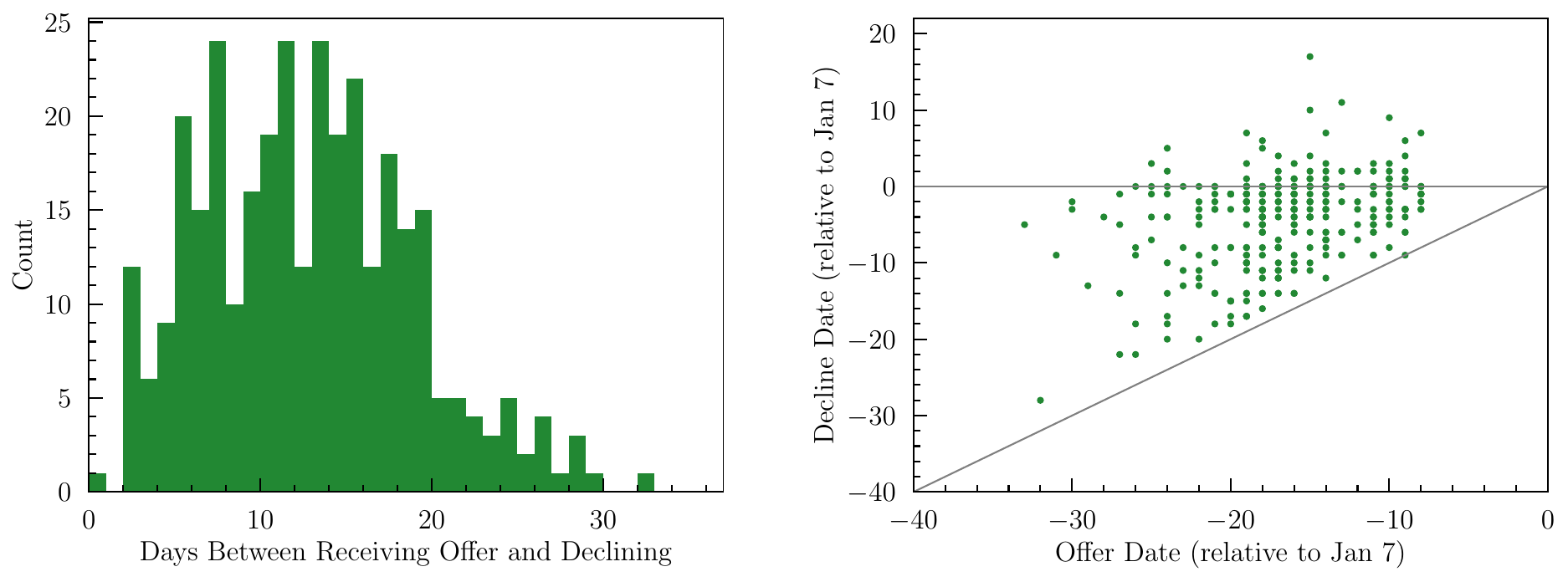}
\caption{\textbf{(left)}~Number of days applicants hold on to offers before declining them. \textbf{(right)}~Correlation between time at which the offer is extended and time at which it is declined. In both plots, only offers extended in the month of December are shown. The sample size for these figures is 326. \label{fig:sitting} }
\end{figure}

One may also wonder how long applicants typically hold on to those offers they end up declining. This distribution is shown in Fig.~\ref{fig:sitting}. This question is however most relevant for \emph{first round} offers, as the consideration time for later offers is typically not determined by the candidates themselves, but by the time remaining to the January 7th deadline. On the other side of the spectrum, there are institutions who  usually extend offers earlier ({\it e.g.}~CERN), and it is expected that those candidates hold on to offers longer as they wait to get a view of their full range of options. To remove these sources of bias, we only consider offers extended in December in Fig.~\ref{fig:sitting}. On average, applicants consider the offers they end up declining for 12 days. Naturally, this time window is getting shorter as the offer is extended closer to the January 7th deadline. The right-hand panel of Fig.~\ref{fig:sitting} shows that a number of offers only get declined \emph{after} January 7th, even if they were extended well in December. We consider it unlikely that all those instances received an extension of the deadline by the institution making the offer. It is more likely that in the majority of those instances the actual posting to the Rumor Mill itself was delayed. 

\clearpage

\section{Community Survey\label{sec:survey}}
To better understand the needs of the community, we conducted a survey through a \href{https://docs.google.com/forms/d/1n236bnEmTBgTtpyrrAangSwvQUPNBPx9DYoJ4At6VFE/edit}{Google form}. The survey was advertised in the following ways: 
\begin{itemize}
\setlength\itemsep{-0.5em}
\item Contributed talk at Theory Frontier P5 meeting on May 4th 2023 \cite{p5}
\item Conference talks: PONT 2023, PLANCK 2023, Pheno 2023 
\item APS division of Particles and Fields newsletter in May 2023
\item Snowmass Theory Frontier mailing list 
\item BSM PANDEMIC mailing list 
\item EUCapt mailing list 
\item UK Cosmo and Hi-Phi (UK HEP) mailing list 
\item Australian mailing list (ARC Dark Matter centre of excellence -- theory group) 
\item Lattice mailing lists: {\tt latticejobs-l@list.indiana.edu, latticenews-l@list.indiana.edu} 
\item Neutrino community: Invisibles mailing list, {\tt neutrinotheory@fnal.gov}, Neutrino theory platform at CERN 
\item Japanese HEP community mailing list
\item Indian HEP community mailing list
\item Former signatories of the 2007 accord 
\item Personal networks and targeted email solicitations.
\end{itemize}
We are still collecting responses and feedback, and invite any additional suggestions for mailing lists and community outreach. 

\subsection{Demographics}

 \begin{table*}[t]
\footnotesize
\begin{center}
\renewcommand{\arraystretch}{1.25}
\begin{tabular}{l|c}
\textbf{Category} & \textbf{Respondents} \\
  \Xhline{3\arrayrulewidth}
Total & 588 \\
\hline
Faculty & 384 \\
Grad \& Postdoc & 204 \\
\hline
North America & 297 \\
Europe & 242 \\
Oceania & 18 \\
Asia & 18 \\
Middle East & 8 \\
Central/South Amer. & 5 \\
\hline 
Pheno & 331 \\
Formal & 179 \\
QCD & 95 \\
Cosmology & 150 \\
Astrophysics & 99 \\
Other & 6\\
\end{tabular}
\end{center}
\caption{\label{tab:numbers} Summary of the distribution of survey respondents.  Note that respondents could select more than one category for their subfield.  The ``Other" category includes gravity and quantum information science.
}
\end{table*}

\begin{figure}\centering
\includegraphics[width=\textwidth]{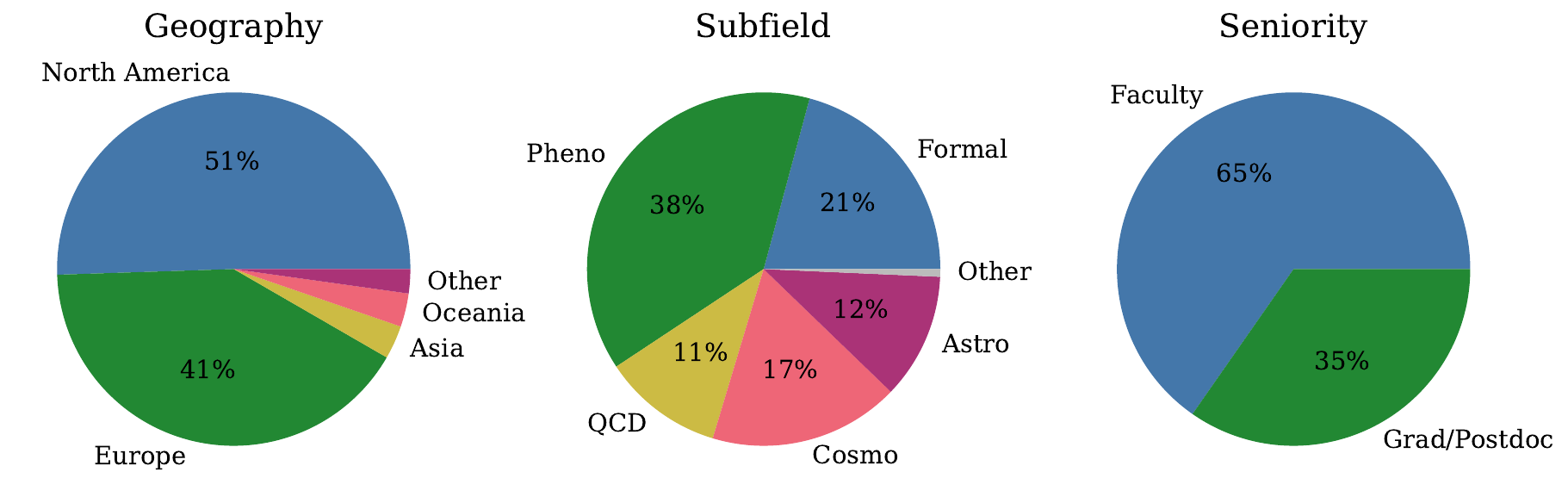}
\caption{Distribution of respondents sorted by several identifiers: \textbf{(left)}~geographical location, where ``other" covers Central and South America, as well as the Middle East; \textbf{(middle)}~self-identified subfield(s), where ``other" covers gravity and quantum information science; and \textbf{(right)} seniority.  Note that respondents could select more than one subfield. \label{fig:demographics} }
\end{figure}

As of June 13, 2023, we have received 588 responses, of whom 530 chose to identify themselves. 
All the named respondents were verified to be members of the HET community.
We can therefore be confident that the survey was not significantly biased by malevolent actors and/or bots. The demographic distribution of the respondents is summarized in Tab.~\ref{tab:numbers} and Fig.~\ref{fig:demographics}. For the geographical information, respondents were asked to indicate where they are currently based as opposed to their region of origin. The options provided were ``North-America,'' ``Europe (including UK),'' ``Asia,'' ``Oceania,'' ``Middle East" and ``Central and South America,'' in addition to a write-in option. For subfield, the given options were ``Formal Theoretical Physics,'' ``Phenomenology/ Beyond the Standard Model,'' ``QCD (including Lattice, SCET, etc.),'' ``Cosmology'' and ``Astrophysics,'' as well as a write-in option. Respondents were allowed to indicate more than one choice on this question, in which case we included them in all categories they indicated. A fraction chose to provide a write-in answer, which were often slight variations on the categories above. In those instances, we included the respondents in the one or more categories considered closest. Finally, for the seniority category, the respondents were asked to chose between ``faculty'' and ``graduate student or postdoc,'' with no write-in option.

We consider the overall distribution of respondents to be reasonably good reflection of the community.  However, it is possible that some sectors of the community did not receive as much advertising as others and that we are thus missing a systematic effect.  To address this concern, the survey will remain open until \textbf{July 21st, 2023} and we hope this arXiv posting (which is cross-listed broadly) will encourage additional responses from anyone who, to date, was unaware of this ongoing exercise. Our results will be updated accordingly, and we will post them on a companion web page~\cite{website}.

\subsection{Results}

\begin{figure}
    \centering
    \includegraphics[width=\textwidth]{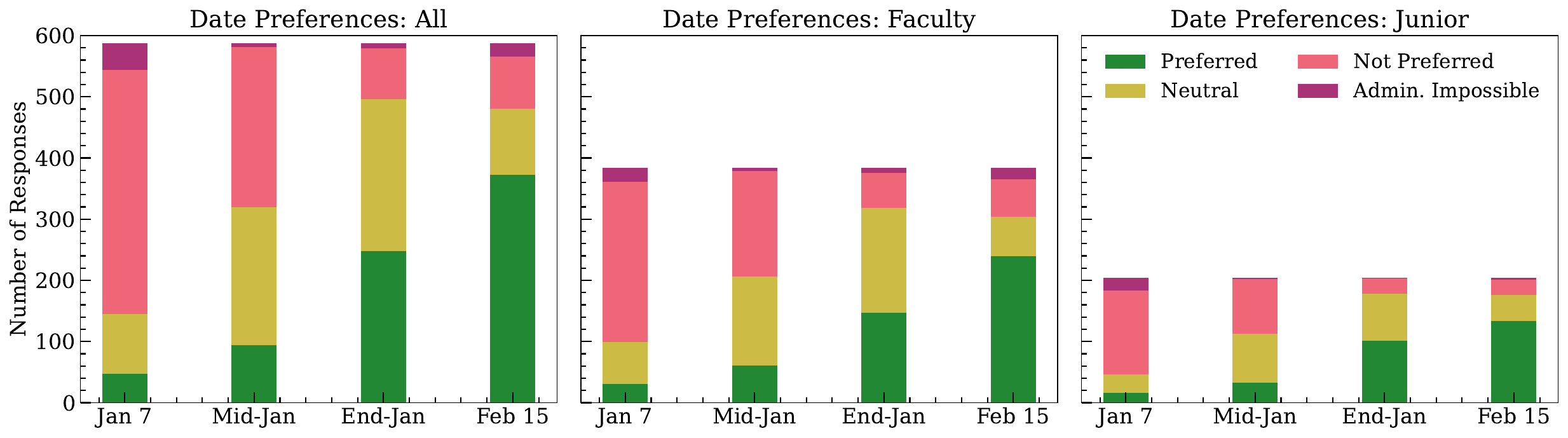}
    \caption{Level of preference among respondents for a January~7, Mid-January, End-of-January, or February~15 acceptance deadline.  Respondents ranked their choice on a scale of: Preferred, Neutral, Not Preferred, and Administratively Impossible.  Results are shown for all respondents~\textbf{(left)}, as well as broken down by faculty~\textbf{(middle)} and junior (grad/postdoc)~\textbf{(right)}.}
    \label{fig:rate}
\end{figure}

 The respondents were asked to rate four options for possible deadline dates:  ``January 7 (current deadline),'' ``Mid-January (around January 15),'' ``End-of-January (around January 30),'' ``February 15 (astronomy deadline).'' These options were chosen after extensive informal interaction across the community,  which indicated little support for bringing the deadline either before the winter break or into March. The results for this question are shown in Fig.~\ref{fig:rate}. A large majority of respondents indicate that the current January 7th deadline is either ``not preferred'' (399 respondents, 68\%) or ``administratively impossible'' (44 respondents, 7\%).
 The degree of preference increases for later dates, with the February 15th option being the most preferred. The ``End-of-January" choice was, by a small margin, the least disliked; it had the smallest contribution of 
``not preferred" (83 respondents, 14\%) and ``adminstratively impossible" (9 respondents, 1.5\%). For the February 15th option, 86 respondents~(15\%) indicated ``not preferred" and 22 respondents~(3.7\%) chose ``adminstratively impossible."
There is no significant difference when breaking the data down according to seniority. These qualitative trends hold true also among the  10\% of respondents that chose to remain anonymous, but with milder hierarchies. 

\begin{figure}
    \centering
    \includegraphics[width=\textwidth]{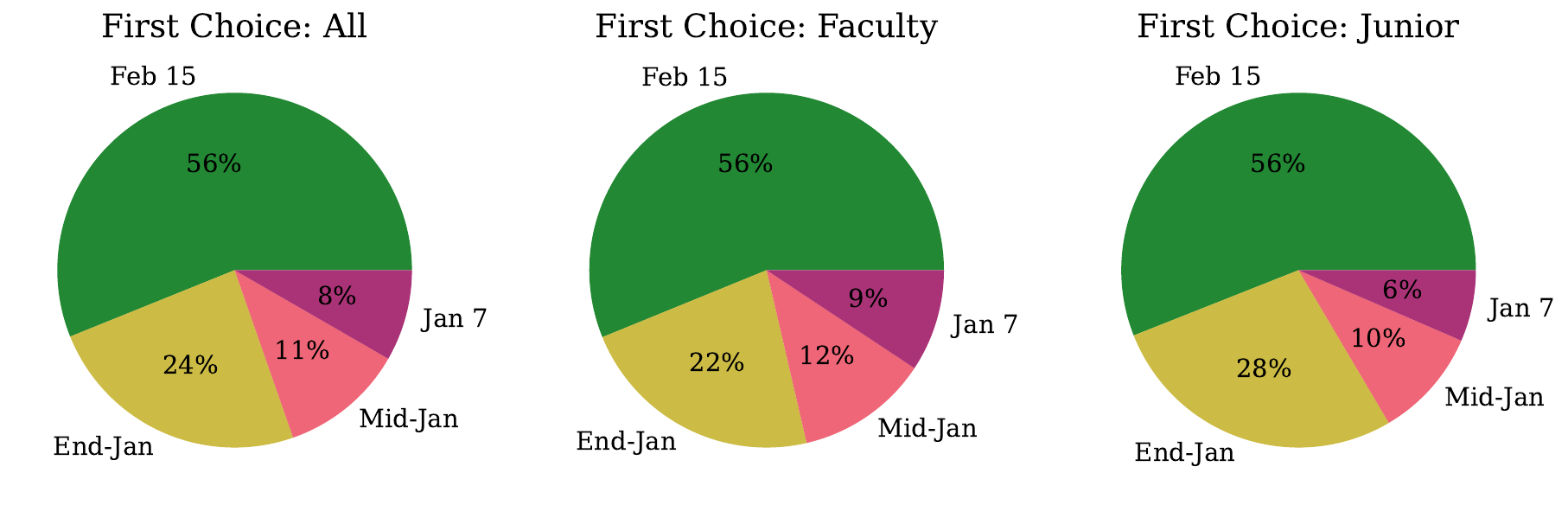}
    \caption{Respondents' first choice for the postdoc acceptance deadline, among the four choices proposed: January~7, Mid-January, End-of-January, and February~15.  Results are shown for all respondents~\textbf{(left)} as well as broken down by  faculty~\textbf{(middle)} and junior (grad/postdoc)~\textbf{(right)}.}
    \label{fig:first_choice}
\end{figure}

\begin{figure}
    \centering
    \includegraphics[width=\textwidth]{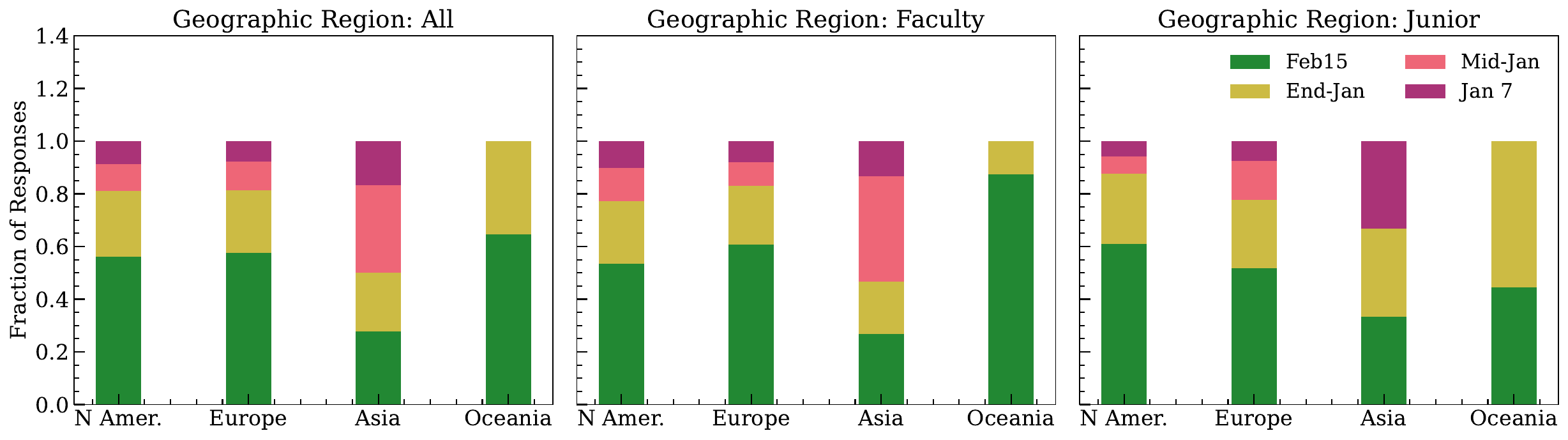}
    \caption{First choice for the acceptance deadline, among the options proposed, broken down by geography and seniority.  The geographic regions correspond to North America~(N Amer.), Europe, Asia, and Oceania.  We caution the reader that the number of respondents in Asia and Oceania is nearly 10\% less than in North America and Europe. Results for the Middle East and Central/South America are not shown given the very small number of respondents from those regions.   }
    \label{fig:first_choice_geo}
\end{figure}

For the same range of options, Fig.~\ref{fig:first_choice} shows the first choice of the respondents for a joint deadline. A majority prefers the mid-February option, followed by the end-of-January and middle-of-January options. The current January 7th deadline is the least preferred among the options proposed, by a rather wide margin. The results are again similar when broken down by  seniority. Fig.~\ref{fig:first_choice_geo} shows the first-choice option, broken down according to geographic region. The two largest groups, Europe and North America are broadly consistent with one another. The responses from Asia are more mixed, and the January~7th deadline appears to enjoy more support than it does in Europe and North America. This may be because of the Lunar New Year, which tends to fall at the end of January or mid-February. There appears to be overwhelming support in Oceania to move away from the early January deadline, likely because it falls in their summer break. For both Oceania and Asia, the sample size is small however, and one should exercise caution in over-interpreting these results. For South/Central America (5 respondents) and the Middle East (8 respondents) the sample size was too small to draw any meaningful conclusions. 

\begin{figure}[p]
    \centering
    \includegraphics[width=0.9\textwidth]{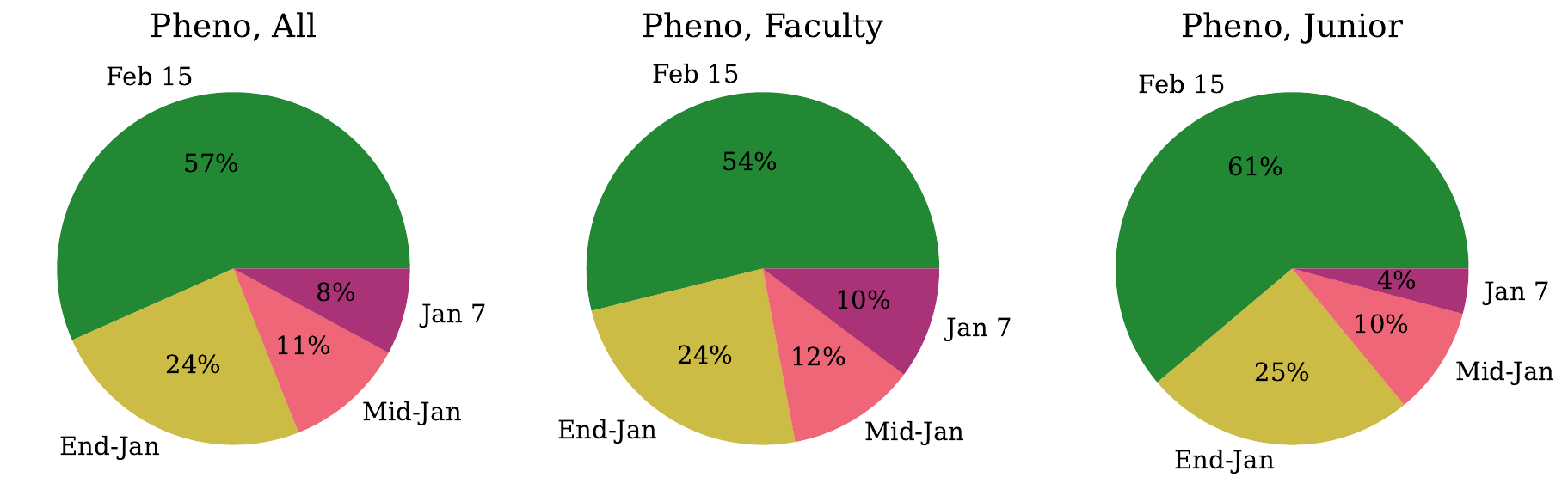}
    \includegraphics[width=0.9\textwidth]{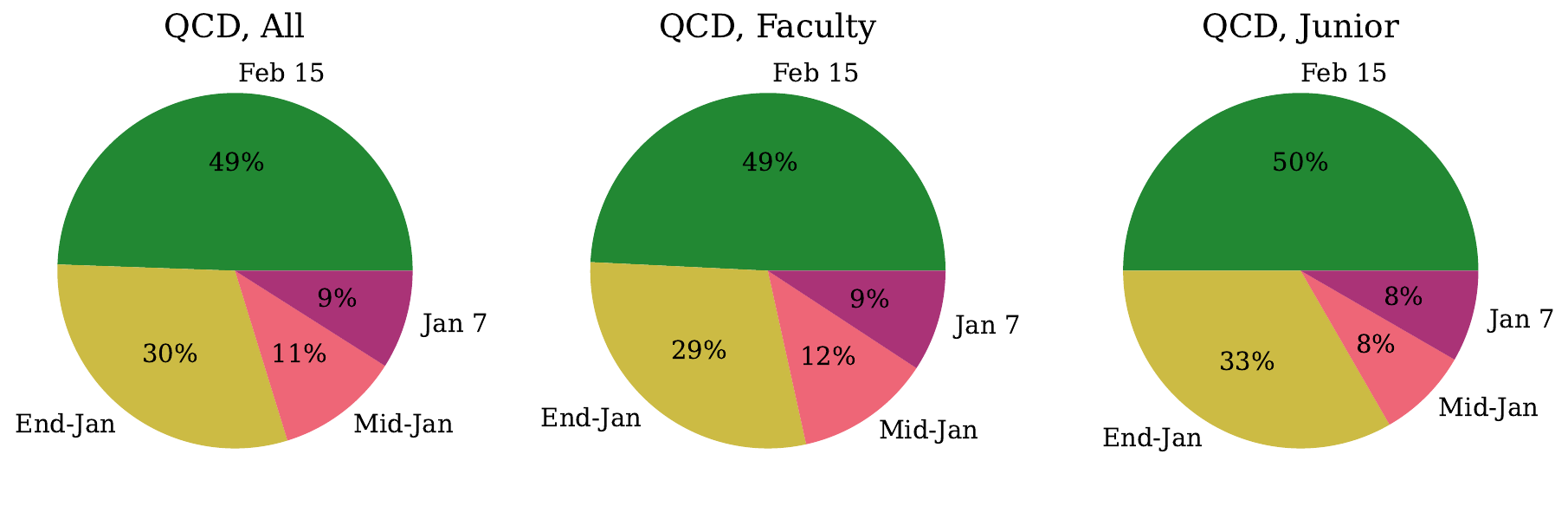}
    \includegraphics[width=0.9\textwidth]{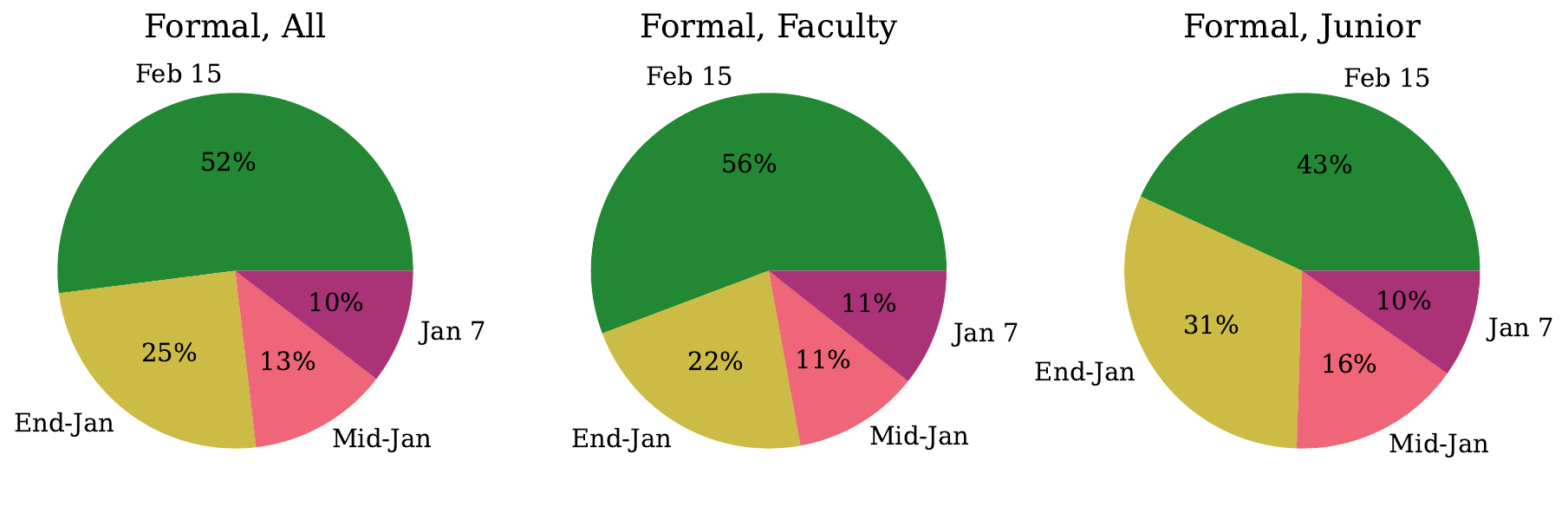}
    \includegraphics[width=0.9\textwidth]{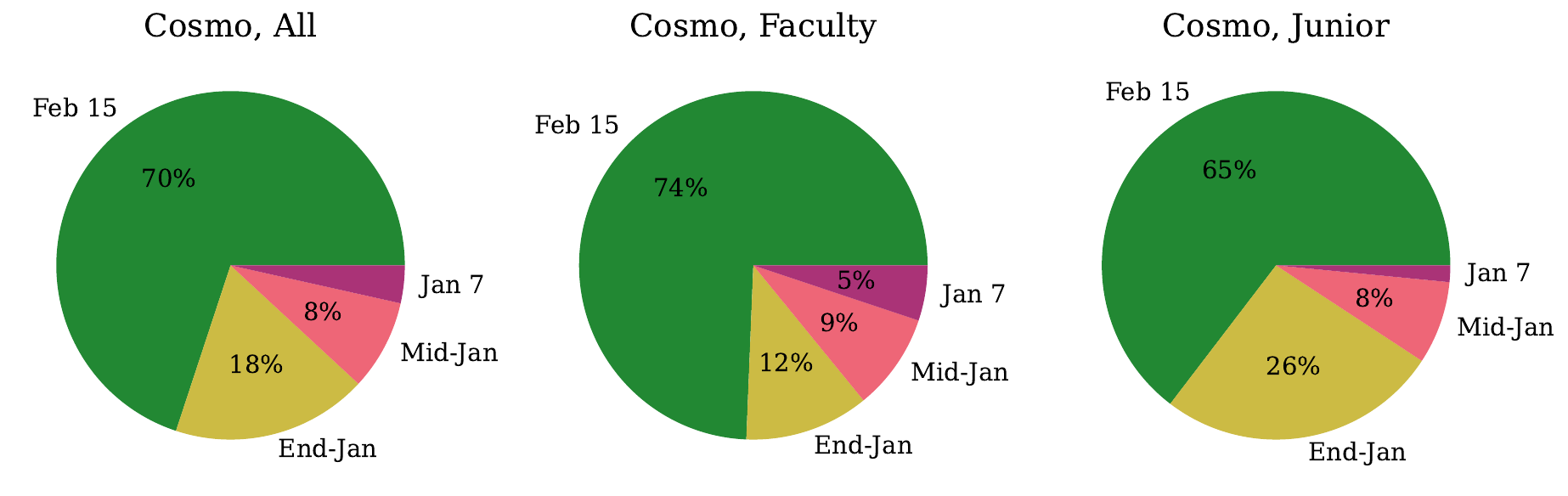}
    \includegraphics[width=0.9\textwidth]{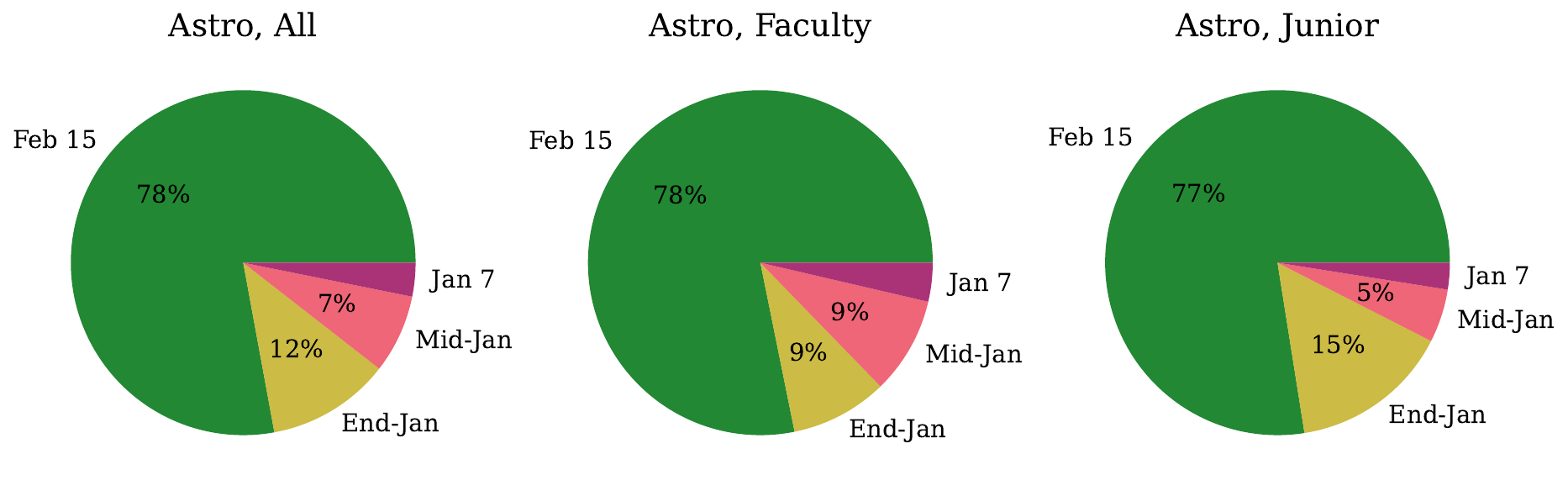}
    \caption{First choice for the acceptance deadline, broken down by subfield and seniority. See Tab.~\ref{tab:numbers} for the number of respondents included in each subfield.\label{fig:prefbysubfield}}
    \label{fig:byfield}
\end{figure}

When broken down according to subfield (see Fig.~\ref{fig:byfield}), the trends described above broadly hold, with one exception. The support for February 15th is much greater among those working on astroparticle physics and/or cosmology. This is not surprising, given the better match with the astrophysics deadline on February 15th. 

\begin{figure}
    \centering
    \includegraphics[width=\textwidth]{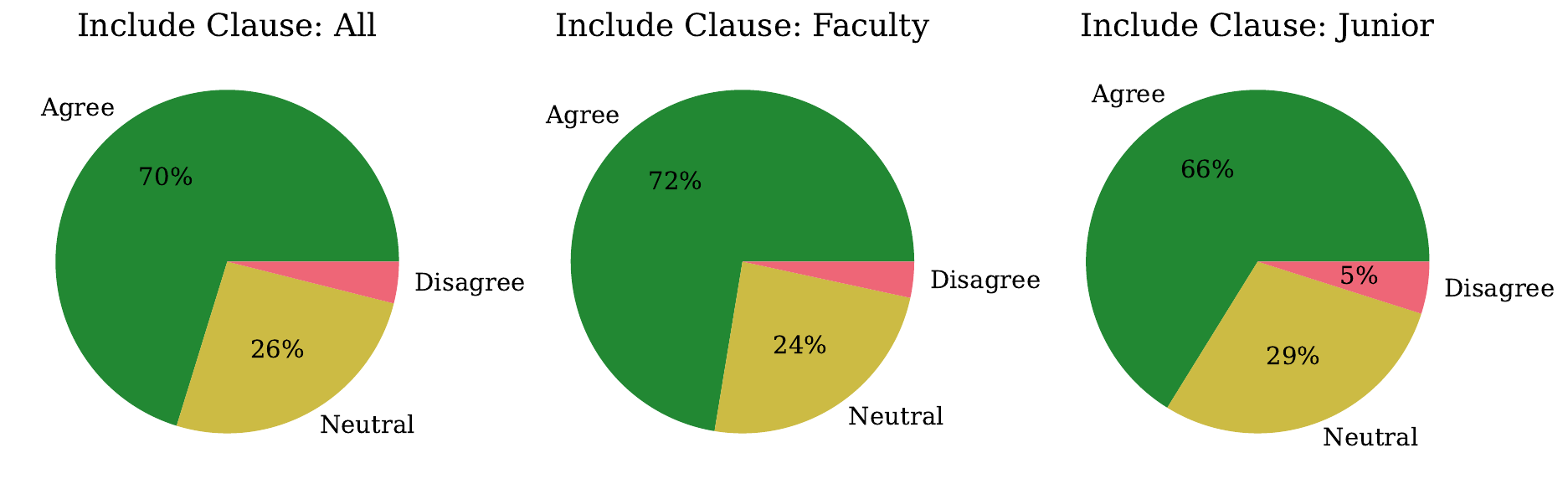}
    \caption{Respondents' opinion as to whether a clause should be included in any new accord that requires the agreement to only go into effect after a large fraction of the original institutions that signed the 2007 accord have signed the new one.   Responses are divided by seniority. }
    \label{fig:clause}
\end{figure}

Finally, it important to note that community unity on the acceptance deadline appears to be of critical importance for many. This was already evident from our many private interactions with colleagues before launching the survey, and the survey questions were therefore deliberately and carefully formulated with this concern in mind. To measure this sentiment to some extent, we asked respondents to indicate whether they agree with the statement: \emph{``If a new agreement is drafted, I would prefer it to include a clause stating that the agreement only go into effect after a large fraction of the institutions which signed the original 2007 agreement have signed on.''} An overwhelming majority responded in the affirmative, as shown in Fig.~\ref{fig:clause}, indicating that the \emph{existence} of a joint deadline outweighs the particular deadline date for a large fraction of the community.

\FloatBarrier

\section{Common Concerns\label{sec:comments}}

We received a total of 138 write-in comments, which we categorised broadly  in the following manner:
\begin{itemize}[itemsep=-\parsep]
    \item Concerns related to the time it takes for visas/work permits to be issued
    \item Dates of fellowship decisions 
    \item Alignment with other fields/subfields
    \item Concerns about CERN
    \item Holidays in early January 
    \item Concern about long gap between application deadlines and decision deadline
    \item The need to have “best practices” 
    \item School/child care registration concerns
    \item Making sure the deadline does not fall on a weekend
    \item General statement of support
\end{itemize}

We note that not all of these categories are decisively positive or negative; for example, we received comments in support of a later deadline because of Epiphany on January 6th, and comments against it because of Lunar New Year. We summarise, with the assistance of \texttt{Croox-NET},
some of the particular issues brought up, without attempting to draw a conclusion from them.

\paragraph{Visa Issues:}

We received 13 comments about various concerns related to visa applications. Particular examples that were highlighted were the H1B visa process in the US and non-EU visa processes in Europe. For example, two respondents deemed it impossible to obtain a work-permit for non-EU citizens after February 15th  in time for a September/October start, due to Belgian immigration rules. 
In addition, some comments expressed the concern that postponing the deadline could result in potential disadvantages for vulnerable groups. Ensuring equal opportunities for candidates from countries without visa waivers is of utmost importance. Moreover, it is crucial to consider the situation of international applicants who do not receive offers in the initial round and consequently face a waiting period before commencing the offer/negotiation process.

\paragraph{Schooling and Family Issues:}
We received 5 comments about issues related to schooling of postdocs' children and other family-related issues. 
It was emphasized that deadlines should be arranged in a way that allows recipients sufficient time to discuss their options with their families and advisors. One important reason for this is the challenge  associated with finding daycare spots for young children. Another highlighted issue is the proximity of the public-school enrollment deadline to the postdoc acceptance deadline. Moreover, many school districts require families to secure housing before allowing enrollment. 
A further reason for this is that partners of applicants also require adequate time to find employment in the new location. As one respondent said, the overarching focus of any postdoc agreement should be on providing flexibility and options to recipients, taking into consideration the profound impact of deadlines on their personal, professional, and family lives.

\paragraph{Holiday Schedules:}
We received 22 comments generally related to how the postdoc hiring season overlaps with the holiday schedule making it hard for faculty to communicate with administrators, postdocs to communicate with mentors, and everyone to be able to have time with their family.
School and daycare closures over the holidays make working over the holidays difficult for people with caring responsibilities.  
It was also noted that the schedule for holidays over the December--February period is very location dependent.  For instance, countries where Epiphany is a publicly celebrated holiday are still on holiday until January 7th, with staff often absent for a little longer.  Institutions in the Southern hemisphere are out-of-sync with their Northern hemisphere counterparts. Additionally, the long summer break is over December--January.
This issue is further complicated by those holiday schedules that are based on the lunar calendar and shift around annually.    
It was also noted by some respondents that {\it application} deadlines should not be shifted back by any substantial amount. 

A related issue that was brought up in 2 comments is that a fixed date for the deadline occasionally falls on a weekend.  This problem could be removed by defining the deadline not as {\it e.g.}~January 30th but instead as the {\it e.g.}~the last Wednesday in January.  However, even this has complications since, for instance, the third Monday in February is a holiday (Presidents' Day) in the US.

\paragraph{Concern about CERN:} We received 4 comments about CERN offers, which typically go out early. A few participants expressed the belief that CERN also imposes a response deadline prior to January 7th, thereby not abiding by the existing accord. This is incorrect, as CERN has always respected the accord. Moreover, the following statement was offered by Gian Giudice on behalf of CERN: 
\begin{quote}
    CERN has an early application deadline for theory postdocs because the selection process requires a first stage done at the individual national level. For this reason, CERN has been making postdoc offers as early as November. However, CERN has {\it always} respected the January 7th response deadline, and intends to respect any new deadline that will be agreed upon by the physics community. CERN welcomes the proposal to shift the January 7th deadline to a later date. In order to adapt better to such changes, CERN is considering to move its current application deadline to a later date, closer to those of other institutions.
\end{quote}

\paragraph{Best Practices:}
We had 6 comments advocating for the development of a set of ``best practices" for postdoc hiring in the field.  These included a \emph{minimum} amount of time that a candidate should be allowed between an offer and a deadline.  This would be especially important for offers made after the deadline in the postdoc accord.  Suggestions of 1 or 2 weeks were made.  It was also suggested that applicants should be strongly encouraged to never hold more than 2 or 3 offers simultaneously, for any extended period, and that  postdoc advisors/mentors should encourage this behavior.  Some respondents furthermore emphasized that Academic Jobs Online~(AJO)~\cite{ajo} provides a central application site and that it would be helpful for applicants as well as letter writers if all institutions utilized it, rather than using their own bespoke interfaces.  Finally, it was pointed out that while interviewing candidates to determine how well-matched they are to an institution is beneficial, candidates should not be asked to make a commitment before receiving an offer.

\paragraph{Fellowship Outcomes:}
A deadline for regular postdoc decisions after the deadlines associated with prestigious fellowships ({\it e.g.}~Marie Skłodowska-Curie Actions) allows applicants with outstanding applications for these fellowships to make decisions with full knowledge of their options. 
It is also beneficial to institutions since it removes the risk of their position being 
vacated shortly 
after acceptance.  This may be of particular importance if the fellowship is cross-disciplinary and the leading HET candidate is not selected because they have accepted a non-fellowship position. Tab.~\ref{tab:fellowships} contains a list of multi-disciplinary fellowships known to us, with their respective deadlines.  We will keep an updated list of multidisciplinary fellowships on our web page~\cite{website} and invite the community to submit such fellowships to us at  \href{mailto:het.postdoc.deadline@gmail.com}{het.postdoc.deadline@gmail.com}.

\begin{table}
\begin{tabular}{l|l}
Fellowship & Acceptance deadline\\\hline
Berkeley Miller &
Typically the third week of January\\
Harvard Society of Junior Fellows &
End of January\\
Marie Skłodowska-Curie Actions&
Offers made mid/late February\\
MIT Pappalardo &
Currently January 7th\\
NASA Hubble &
February 15th\\
Princeton PCTS&
No official acceptance deadline; \\
& informal encouragement to respond by early January\\
UC Presidential and Chancellor&
Offers at the end of February\\
Los Alamos fellowship & Offers early February\\
\end{tabular}
\caption{List of special fellowships, known to us, and their deadlines. The current January 7th deadline for the Pappalardo fellowship is specifically set to align with the existing HET deadline.
\label{tab:fellowships} }
\end{table}

\paragraph{Alignment:} 
There were 20 comments about alignment with other deadlines. Several respondents identified the need for alignment between postdoc application deadlines and the timing of permanent job/research assistant professor/fellowship searches to prevent star candidates from turning down multiple postdoc offers for permanent positions. Others focused on the coordination of deadlines between fields---interdisciplinary applicants working in both particle physics and either astrophysics or mathematics. Maximising alignment 
favors the deadlines in February over those in January. 

\section{Conclusion\label{sec:conclusions}}
We conducted a community survey regarding the desirability of  January 7th as the common postdoc offer acceptance deadline. We will continue to collect responses on our \href{https://docs.google.com/forms/d/1n236bnEmTBgTtpyrrAangSwvQUPNBPx9DYoJ4At6VFE/edit}{survey} until \textbf{July 21st, 2023} and will update this note if we observe a significant shift in the results.

As of June 13, 2023, 588 physicists have participated in the survey, broadly representing the various subfields, geographical areas and levels of seniority present in the HET community. The January 7th deadline is broadly disliked and later options such as end-of-January or mid-February gather a significantly greater degree of support. In addition, the later options are in-line with the established deadlines of the mathematics and astrophysics communities. Community unity on a common deadline is widely considered to be very important, both among respondents to the survey and from private conversations with colleagues. Moreover, a number of important concerns were brought up, both with respect to the existing January 7th deadline as well as with potential alternatives.

We believe that the results of this survey, as they are in July 2023, indicate that a revision of the January 7th common deadline is favored by the community. That is, provided that most concerns raised can be addressed adequately, while keeping in mind that no choice will be able to alleviate all concerns perfectly, including the status quo. In the next few months, we therefore aim to draft an updated accord, compile answers to frequently asked questions and compile a list of ``best practices'' for the postdoc hiring process, which will be independent of the accord itself. Our aim is to have a new accord completed and ratified before the Fall 2023 application cycle. They will appear on the following webpage:

\begin{center}
\url{https://het-postdoc-accord.github.io/information/}
\end{center}
The website contains frequently asked questions and a periodically-updated list of supporters. 
In the meanwhile, we welcome further responses to the \href{https://docs.google.com/forms/d/1n236bnEmTBgTtpyrrAangSwvQUPNBPx9DYoJ4At6VFE/edit}{survey} until \textbf{July 21st, 2023} from those who have not yet responded.  We also continue to welcome further comments, concerns, and suggestions, which can be sent to  \href{mailto:het.postdoc.deadline@gmail.com}{het.postdoc.deadline@gmail.com}.

\section*{Acknowledgments}
We are grateful to Felix Yu for providing us with the raw data from the Postdoc Rumor Mill. We thank the many colleagues who gave us informal feedback on the deadline issue itself, how to proceed prudently and on the design of the survey. We are particularly grateful to Robert McGehee, Nathaniel Craig, Gian Giudice, Joachim Kopp, Rachel Houtz,  and Zhen~Liu  for providing exceptionally detailed input and/or for bringing additional subtleties to our attention.

\bibliographystyle{jhep}
\bibliography{refs}

\end{document}